\documentclass[journal]{IEEEtran}
\ifdefined\pdfobjcompresslevel\pdfobjcompresslevel=0\fi
\ifdefined\pdfcompresslevel\pdfcompresslevel=9\fi

\usepackage[T1]{fontenc}
\usepackage[utf8]{inputenc}
\usepackage{graphicx}
\usepackage{cite}
\usepackage{url}
\usepackage{float}
\usepackage{amsmath}
\usepackage{amssymb}
\usepackage{pgfplots}
\pgfplotsset{compat=newest}
\usepackage[caption=false,font=footnotesize]{subfig}
\usepackage{tikz}
\usepackage{tabularx}
\usepackage{booktabs}
\usepackage{array}
\usepackage{hyphenat}
\usepackage{siunitx}
\usepackage{enumitem}
\usepackage{dblfloatfix}
\usepackage{placeins}
\usepackage{needspace}
\usepackage{xspace}
\usetikzlibrary{arrows.meta, positioning}

\DeclareSIUnit{\molar}{M}
\DeclareSIUnit{\micrometerunit}{\ensuremath{\mu}\mathrm{m}}
\newcolumntype{Y}{>{\raggedright\arraybackslash}X}
\graphicspath{{Figures/}}

\DeclareGraphicsExtensions{.pdf,.png,.jpg,.jpeg}


\newcommand{\DA}{\ensuremath{\mathrm{DA}}\xspace}
\newcommand{\HT}{\ensuremath{\mathrm{5\text{-}HT}}\xspace}
\newcommand{\CTRL}{\ensuremath{\mathrm{CTRL}}\xspace}

\newcommand{\frameworkname}{MC-BRIDGE\xspace}
\newcommand{\frameworkfull}{Molecular Communication Bioelectronic Receiver-chain Integrated Design and Guided Evaluation\xspace}

\newcommand{\biophoto}[1]{\IfFileExists{#1}{\includegraphics[width=0.82in,height=1.025in,clip,keepaspectratio]{#1}}{\fbox{\parbox[c][1.025in][c]{0.82in}{\centering Photo}}}}

\makeatletter
\def\@IEEEBIOphotowidth{0.82in}
\def\@IEEEBIOphotodepth{1.025in}
\def\@IEEEBIOhangwidth{0.96in}
\def\@IEEEBIOhangdepth{1.025in}
\def\@IEEEBIOskipN{1.5\baselineskip}
\makeatother
\newif\ifincludeauthorbios
\includeauthorbiostrue

\title{\frameworkname: A Modular Receiver-Chain Simulation Framework for OECT-Based Molecular Communication}
\author{Hongbin Ni, \textit{Student Member, IEEE} and Ozgur B. Akan, \textit{Fellow, IEEE}%
\thanks{The authors are with the Centre for neXt Communications (CXC), Electrical Engineering Division, Department of Engineering, University of Cambridge, CB3 0FA Cambridge, U.K. (e-mail: hn345@cam.ac.uk).}%
\thanks{Ozgur B. Akan is also with the Center for neXt Communications (CXC), Department of Electrical and Electronics Engineering, Ko\c{c} University, 34450 Istanbul, T\"{u}rkiye (e-mail: oba21@cam.ac.uk, akan@ku.edu.tr).}%
}

\begin{document}
\flushbottom

\maketitle

\begin{abstract}
Organic electrochemical transistor (OECT)-based molecular communication (MC) receivers connect transport and binding to device current, noise, calibration, and detection. We present \frameworkname (\frameworkfull), a modular framework linking release, extracellular diffusion, finite-area observation, stochastic binding, and OECT transduction to charge-domain detection. On an independent electrical time grid, it generates sequence-wide multichannel colored noise, performs control-channel referencing and charge integration, and supports molecule-shift-keying (MoSK), concentration-shift-keying (CSK), and Hybrid decisions. Through common module interfaces, it estimates symbol error rate (SER) and decoded-symbol mutual information and analyzes inter-symbol interference (ISI). At nominal separation, the passive finite-area observer retains $27.5\%$ of the center-point decision-charge magnitude, while selective--control correlation determines whether control referencing helps. After adaptive search, held-out records with seeds disjoint from search and calibration test the selected and next-lower budgets, yielding a tested-grid upper bound on the minimum budget meeting the SER target. This passive-field test resets receptor occupancy, excludes ISI, and calibrates thresholds on separate records at each operating point. Retaining transport and receptor memory instead produces high SER. Thus, geometry, covariance, calibration, and memory can change receiver conclusions.
\end{abstract}


\section{Introduction}
\label{sec:introduction_related_scope}

\IEEEPARstart{M}{olecular} communication (MC) uses particles as information carriers and provides a communication-theoretic framework for nanoscale, microscale, and biological information exchange \cite{akyildiz_iobnt_2015,farsad_survey_2016,kuscu_survey_2019}. A demanding setting is organoid-facing molecular readout. Brain organoids are self-organized three-dimensional neural tissues derived from pluripotent stem cells \cite{lancaster_organoids_2013} that can exhibit structured oscillatory activity over long culture times \cite{trujillo_organoids_2019}. Their state is shaped not only by electrophysiology but also by extracellular chemical transport, neuromodulator release, uptake, and receptor-level interactions. An MC receiver for this environment must connect molecule budget, symbol timing, and detector error to effective extracellular diffusion, binding kinetics, transconductance, drift, and correlated readout noise. Because geometry, covariance, calibration, and memory couple across layers, controlled comparison of complete OECT-MC receiver designs requires an end-to-end framework in addition to isolated component models.

Organic electrochemical transistors (OECTs) convert ionic and chemical perturbations into electronic current through volumetric mixed ionic--electronic conduction \cite{rivnay_oect_review_2018,friedlein_device_physics_2018}. High-transconductance poly(3,4-ethylenedioxythiophene):polystyrene sulfonate (PEDOT:PSS) devices and on-site OECT amplification for electrochemical aptamer sensing have been demonstrated \cite{khodagholy_high_transconductance_2013,ji_oect_aptamer_2023}. Aptamer-functionalized field-effect devices have also sensed dopamine and serotonin under physiological ionic strength \cite{nakatsuka_aptamer_fet_2018,gao_multiplexed_2022}. These results motivate OECT-based MC receivers in which concentration is an intermediate state rather than the final observable. In \frameworkname (\frameworkfull), calibrated charge statistics from OECT currents feed molecule-shift keying (MoSK), concentration-shift keying (CSK), or Hybrid decisions, with matched-control subtraction applied to amplitude branches when enabled. Decisions reflect finite release, effective extracellular diffusion, clearance and inter-symbol interference (ISI), stochastic occupancy, OECT transduction, correlated electrical noise, reference processing, and calibration.

Receiver-centric biological field-effect transistor (BioFET) models have already coupled transport, ligand--receptor kinetics, electrical transduction, device noise, and receiver metrics. Kuscu and Akan derive an end-to-end silicon-nanowire field-effect transistor (SiNW-FET) receiver and symbol-error-probability analysis \cite{kuscu_sinw_receiver_2016}. Abdali and Kuscu formulate a frequency-domain microfluidic graphene BioFET channel verified against particle simulations \cite{abdali_graphene_biofet_2024}, while Civas \emph{et al.} study cross-reactive binding-noise detection with particle-based verification \cite{civas_cross_reactive_2024}. \frameworkname builds on this lineage with a common OECT integrated-charge statistic, replaceable modules with explicit input and output quantities, multichannel colored noise generated continuously over the full sequence, general cross-channel correlation matrices for each noise component, separate molecular and electrical time grids, calibration-transfer tests, independent evaluation runs, and automated numerical checks. These features allow candidate receiver designs to be compared using the same OECT charge statistic while retaining intermediate physical states and receiver diagnostics.

A companion receiver-centric study introduces and analyzes a control-referenced tri-channel OECT front end with dopamine (DA)-selective, serotonin (5-HT)-selective, and hydrogel-matched control (CTRL) pixels for Hybrid MC toward brain organoid interfaces \cite{ni_tri_channel_oect_arxiv_2026}. That study asks when the tri-channel architecture benefits Hybrid decoding. The present paper uses the shared model family as a reference scenario to establish a reusable computational method for evaluating candidate OECT-MC receiver configurations. Table~\ref{tab:companion_delta} distinguishes the reused physical-layer model family from the present framework, verification, and inference contributions. The DA/5-HT/CTRL receiver provides a demanding benchmark scenario. Here, we replace the observer and binding implementations, verify full-sequence noise and covariance, and evaluate calibration transfer, sensitivity to memory depth, and the held-out budget bound. Cross-study numerical comparisons require matched geometry, observer, binding update rule, noise, calibration, ISI, and Monte Carlo protocols.

\begin{table*}[!t]
\caption[Relationship to the companion receiver study]{Relationship between the companion receiver study and the present methodological contribution.}
\label{tab:companion_delta}
\centering
\scriptsize
\renewcommand{\arraystretch}{1.02}
\setlength{\tabcolsep}{3.2pt}
\begin{tabularx}{\textwidth}{p{0.16\textwidth} p{0.25\textwidth} p{0.31\textwidth} X}
\toprule
\textbf{Element} & \textbf{Companion receiver study} & \textbf{Present paper} & \textbf{Reuse or new evidence} \\
\midrule
Research question & When the DA/5-HT/\CTRL architecture improves Hybrid reception & How an OECT-MC receiver configuration is specified, verified, and compared & New methodological objective \\
Reference model & Restricted diffusion, Langmuir binding, quasi-static OECT mapping, noise, and MoSK/CSK/Hybrid decisions & One demanding scenario within the framework & Reused and identified as such \\
Computational contribution & Receiver-specific Monte Carlo evaluation and architecture sweeps & Explicit module input/output specifications, general component correlation matrices, an independent electrical grid, and colored-noise synthesis over the full sequence & New framework architecture and verification method \\
Verification and inference & Receiver checks and Monte Carlo schedules set by confidence criteria & Independent Brownian and stochastic simulation algorithm (SSA) benchmarks, analytical-to-sampled covariance agreement, observer convergence, block bootstrap, and independent SER evaluation & New verification evidence \\
Demonstrations & Control regimes, modes, molecule-budget thresholds, timing, and device co-design & Finite-area and binding comparisons, configuration reuse, calibration transfer, a controlled \CTRL-correlation sweep, and sensitivity to ISI and memory depth & New methodological evidence \\
\bottomrule
\end{tabularx}
\end{table*}

\subsection{Related Simulators and Remaining Gap}
\label{subsec:related_simulators_gap}

MC network simulators span Brownian propagation, absorbing reception, modulation, ISI, and standards-aligned end-to-end studies \cite{ieee1906_2016,gul_nanons_2010,llatser_n3sim_2014,yilmaz_mucin_2014,cevallos_ieee1906_ns3_2026}. Reaction--diffusion and neural tools provide spatial microphysics or neural and synaptic observables \cite{noel_accord_2017,andrews_smoldyn_2010,stiles_mcell_2001,hines_neuron_1997,turgut_n4sim_2022}. Table~\ref{tab:simulator_comparison} positions these complementary tools and receiver-centric BioFET models relative to the OECT charge-domain workflow.

\begin{table*}[!t]
\caption[Related framework positioning]{Comparison of representative MC simulation and receiver-model families with the scope of \frameworkname.}
\label{tab:simulator_comparison}
\centering
\scriptsize
\renewcommand{\arraystretch}{1.0}
\setlength{\tabcolsep}{4pt}
\begin{tabularx}{\textwidth}{p{0.23\textwidth} p{0.31\textwidth} X}
\toprule
\textbf{Framework family} & \textbf{Primary scope} & \textbf{Relation to \frameworkname} \\
\midrule
NanoNS, N3Sim, MUCIN, and \mbox{ns-3-based} MC simulation \cite{gul_nanons_2010,llatser_n3sim_2014,yilmaz_mucin_2014,cevallos_ieee1906_ns3_2026} & MC network abstraction, Brownian propagation, absorbing reception, modulation, and ISI & Transport and communication outputs can feed receiver-facing binding, OECT, covariance, and calibration layers. \\
AcCoRD, Smoldyn, MCell, NEURON, and N$^4$Sim \cite{noel_accord_2017,andrews_smoldyn_2010,stiles_mcell_2001,hines_neuron_1997,turgut_n4sim_2022} & Spatial reaction--diffusion, neural, or synaptic simulation & Their physical or neural observables are complementary inputs to an organoid-facing OECT receiver analysis. \\
Receiver-centric BioFET models \cite{kuscu_sinw_receiver_2016,abdali_graphene_biofet_2024,civas_cross_reactive_2024} & End-to-end SiNW/graphene BioFET signal, noise, and detection models, including frequency-domain and cross-reactive reception & They establish receiver integration and analyses verified against particle simulations. \frameworkname emphasizes reusable OECT charge-domain interfaces, multichannel covariance, calibration transfer, independent evaluation, and automated numerical checks. \\
\frameworkname & Modular OECT-MC analysis from release to decoded symbol & Finite-area observation, stochastic occupancy, OECT readout, full-sequence colored noise and cross-channel covariance, calibrated decisions, an independent tested-grid molecule-budget bound, and decoded-symbol information diagnostics. \\
\bottomrule
\end{tabularx}
\end{table*}

\subsection{Scope and Contributions}
\label{subsec:scope_contributions}

The contributions of this paper are as follows:
\begin{enumerate}[leftmargin=*,label=\arabic*)]
  \item \textit{Modular OECT-MC receiver framework:} We present a modular receiver simulation framework linking release, transport, binding, OECT transduction, electrical noise, control referencing, and detection through interfaces that preserve the physical and statistical meaning of quantities passed between layers.
  \item \textit{Consistent electrical and statistical treatment:} We synthesize multichannel noise continuously over the full sequence on an electrical grid independent of the molecular step, use a common one-sided power spectral density (PSD) convention for sampled and analytical charge covariance, accept general positive-semidefinite component correlation matrices, and use independent evaluation data to support a molecule-budget bound on the tested grid.
  \item \textit{Verification and transfer evidence:} We compare analytical transport with Brownian simulations, exact-transition binding with Gillespie SSA, center-point with square-area observation, and lumped with nodewise binding. We also test a signal-matched low-occupancy case, calibration transfer, sweeps of selective--\CTRL correlation with DA--5-HT correlation held fixed, and sensitivity to ISI and memory depth.
\end{enumerate}

\noindent\textit{Code availability:} The MC-BRIDGE source code is available at \url{https://github.com/athan614/MC-BRIDGE_Simulator}.

\section{\frameworkname Cross-Layer Receiver Framework and Module Interfaces}
\label{sec:architecture}

\begin{figure*}[!tbp]
  \centering
  \includegraphics[width=0.96\textwidth]{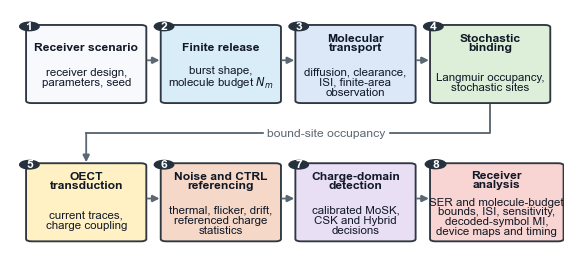}
  \caption[Simulator architecture]{\frameworkname receiver chain. Scenario parameters define the candidate receiver configuration. The baseline path maps finite release to effective extracellular diffusion, stochastic binding, OECT transduction, full-sequence electrical noise, \CTRL referencing, and charge-domain detection. The final stage reports receiver metrics and numerical-verification results.}
  \label{fig:simulator_architecture}
\end{figure*}

\frameworkname maps a candidate receiver configuration through release, transport, binding, transduction, noise, matched-control processing, and charge-domain detection. Fig.~\ref{fig:simulator_architecture} shows the receiver chain and module interfaces.

\subsection{Receiver-Chain Abstraction}
\label{subsec:signal_flow}

A transmitted symbol sequence passes through the following replaceable modules,
\begin{equation}
  b_i \xrightarrow{\mathcal{R}} s_i(t) \xrightarrow{\mathcal{D}} \mathbf{c}_i(t)
  \xrightarrow{\mathcal{B}} \mathbf{N}_{b,i}(t) \xrightarrow{\mathcal{G}}
  \mathbf{i}_i(t) \xrightarrow{\mathcal{N}} \mathbf{y}_i(t)
  \xrightarrow{\mathcal{Q},\mathcal{H}} \widehat{b}_i,
  \label{eq:module_chain}
\end{equation}
where $\mathcal{R}$ is release, $\mathcal{D}$ is transport and ISI, $\mathcal{B}$ is binding, $\mathcal{G}$ is OECT transduction, $\mathcal{N}$ is noise, $\mathcal{Q}$ forms windowed charges, and $\mathcal{H}$ is the detector. Vector interfaces support configurations with one, two, or three channels. In the reference case, CTRL carries no intended analyte-binding signal and enters amplitude-reference processing.

A study configuration fixes the source, geometry, device, covariance, window, modulation, calibration, budget grid, and random streams so that comparisons differ only in the specified factors.

\subsection{Module Interfaces and Alternative Models}
\label{subsec:module_contracts}

We use module contract to mean the input/output specification that identifies the physical quantity, units, dimensions, and metric definitions shared by compatible implementations. Table~\ref{tab:module_contracts} gives this specification, the baseline implementation, and representative alternatives for each module. Every alternative retains the common interface and adds a component-level numerical check.

\begin{table*}[!t]
\caption[Module interfaces and alternative models]{Module interfaces in \frameworkname. Alternative models preserve the stated input/output quantity and include a component-level check.}
\label{tab:module_contracts}
\centering
\scriptsize
\renewcommand{\arraystretch}{1.00}
\begin{tabularx}{\textwidth}{p{0.13\textwidth} p{0.24\textwidth} p{0.25\textwidth} X}
\toprule
\textbf{Module} & \textbf{Baseline implementation} & \textbf{Input/output specification} & \textbf{Representative alternative models} \\
\midrule
Scenario/configuration & Seeds, timing, budgets, channel parameters, and analyses & Candidate receiver design $\rightarrow$ resolved run settings & Calibration import, experiment templates, measured configurations \\
Release & Finite burst with optional stochastic count & Symbol and budget $\rightarrow s_i(t)$ & Vesicular, gamma, or measured waveforms \\
Transport/ISI & Effective diffusion, point or square observer, clearance, and finite memory & Release and geometry $\rightarrow$ concentration traces or window coefficients & Particle transport, hydrogels, finite boundaries, advection \\
Binding & Langmuir mean and stochastic birth--death occupancy, with no specific binding on \CTRL & Local concentration $\rightarrow N_b(t)$ & Cross-reactivity, nonspecific or cooperative binding \\
OECT transduction & Quasi-static $g_m$, $C_{\mathrm{tot}}$, and signed charge coupling & Bound sites $\rightarrow$ noiseless current & Dynamic or measured transfer response \\
Noise/reference & Sequence thermal, flicker, and drift with component covariance and \CTRL subtraction & Noiseless currents $\rightarrow$ noisy or referenced charges & Measured spectra, nonstationary drift, adaptive referencing \\
Detection/calibration & MoSK, CSK, Hybrid, disjoint-seed calibration, and windowed charges & Channel statistics $\rightarrow$ decisions and diagnostics & Sequence detection, online recalibration, equalization \\
Receiver analysis & SER, tested-grid molecule-budget evaluation, ISI, sensitivity, timing, decision surfaces, and empirical mutual information & Decisions, traces, statistics, and settings $\rightarrow$ receiver metrics and numerical results & Latency, energy, utility, measured-data scenarios \\
\bottomrule
\end{tabularx}
\end{table*}

The case studies replace center-point observation with finite-area averaging, compare spatially lumped binding with binding solved at each quadrature node, and reuse the receiver chain for a single-axis CSK-4 receiver. The three substitutions are evaluated using quadrature convergence, decision-charge agreement, and common receiver metrics, respectively. The same comparisons are repeated for a signal-matched low-occupancy case.

\section{Baseline Model and Numerical Verification Protocol}
\label{sec:baseline_model_validation}

\subsection{Release, Effective Transport, and ISI}
\label{subsec:baseline_transport}

For symbol interval $i$, the transmitter selects a molecule species $m_i \in \mathcal{M}=\{\DA,\HT\}$ and an amplitude level $a_i$ determined by the modulation format. Symbols are independent and equiprobable in the reported benchmarks. MoSK uses $a_i=1$. For $M$-ary CSK, level $\ell\in\{0,\ldots,M-1\}$ uses $a_{\ell}=2(\ell+1)/(M+1)$, so the average amplitude is one. CSK-4 therefore uses $\{0.4,0.8,1.2,1.6\}N_m$. Hybrid uses low and high levels $\{2/3,4/3\}N_m$ on each species, with symbols 0 and 1 assigned to DA and symbols 2 and 3 assigned to 5-HT. Thus, $N_m$ is the mean nominal release per equiprobable symbol. The nominal molecule budget is $\bar{N}_{m,i}=a_i N_m$. When molecular shot noise is enabled, the realized emission count is
\begin{equation}
  \widetilde{N}_{m,i} \sim \operatorname{Poisson}(\bar{N}_{m,i}),
  \label{eq:poisson_release_count}
\end{equation}
with zero emission for nonselected species. The baseline release waveform is a finite rectangular burst,
\begin{equation}
  s_i^{(m)}(t)=\frac{\widetilde{N}_{m,i}}{T_{\mathrm{rel}}}
  \mathbf{1}_{0\leq t-t_i<T_{\mathrm{rel}}},\qquad t_i=(i-1)T_s,
  \label{eq:rectangular_release}
\end{equation}
where $T_s$ is the symbol period and $T_{\mathrm{rel}} \ll T_s$ is the release duration. The release module also supports alternative normalized source profiles, but the rectangular burst is used for the baseline model and case studies.

Transport is modeled with a three-dimensional effective extracellular-medium Green's function for a point source at normal separation $r_X$ from receiver channel $X$, with extracellular volume fraction $\alpha$, tortuosity $\lambda$, and first-order clearance $k_{\mathrm{clear}}$ \cite{nicholson_extracellular_1998,sykova_diffusion_2008}. For species $m$,
\begin{equation}
  G_m(r_X,t)=
  \frac{\mathbf{1}_{t>0}}{\alpha(4\pi D_{\mathrm{eff},m}t)^{3/2}}
  \exp\!\left[-\frac{r_X^2}{4D_{\mathrm{eff},m}t}-k_{\mathrm{clear}}t\right],
  \label{eq:restricted_green}
\end{equation}
where $D_{\mathrm{eff},m}=D_m/\lambda^2$. The corresponding local molar concentration is
\begin{equation}
  c_m^{(X)}(t)=\frac{1}{1000N_A}\int_{0}^{t}s^{(m)}(\tau)G_m(r_X,t-\tau)d\tau,
  \label{eq:transport_convolution}
\end{equation}
where $N_A$ is Avogadro's constant and the factor $1000$ converts molecules per cubic meter to molar units. For the rectangular release used in the case studies, the convolution in \eqref{eq:transport_convolution} is evaluated analytically for a finite burst. Independent quadrature provides a numerical reference. Because the \SI{200}{\micrometerunit} gate side is not small relative to the reference separations, the baseline model uses the area-averaged passive observer
\begin{equation}
  \overline{c}_m^{(X)}(t)=\frac{1}{L_g^2}
  \int_{-L_g/2}^{L_g/2}\!\int_{-L_g/2}^{L_g/2}
  c_m\!\left(\sqrt{r_X^2+x^2+y^2},t\right)dxdy,
  \label{eq:square_area_observer}
\end{equation}
where $L_g$ is the square-gate side length. Tensor Gauss--Legendre quadrature evaluates \eqref{eq:square_area_observer}. The center-point observer is retained as the $L_g\rightarrow0$ limiting case and for comparison with the square-area implementation. To assess spatial lumping, we also solve the Langmuir model independently at each quadrature radius and integrate the bound-state response over the gate. In what follows, $c_{m,\mathrm{obs}}^{(X)}(t)$ denotes the selected observer output, which is either $c_m^{(X)}(t)$ for the point implementation or $\overline{c}_m^{(X)}(t)$ for the square-area implementation.

The separation $r_X$ is an effective normal source-to-channel distance. Equal channel distances define an idealized symmetric geometry for isolating receiver processing and matched-reference effects. Physical layouts can assign channel-specific distances.

For ISI-enabled simulations, concentration contributions from previous symbols are superposed. Let $W=\eta T_s$, with $\eta\in(0,1]$, denote the tail-anchored decision-window length. The window-averaged contribution of a symbol emitted $k$ intervals earlier can be written as
\begin{equation}
  h_{k,m}^{(X)}=\frac{1}{W}\int_{T_s-W}^{T_s}
  \frac{1}{1000N_A}\left[p_{\mathrm{rel}}*\overline{G}_m^{(X)}\right](t+kT_s)dt,
  \label{eq:window_coefficients}
\end{equation}
where $p_{\mathrm{rel}}(t)$ is the unit-area release profile and $\overline{G}_m^{(X)}$ denotes the Green's function after applying the selected point or square-area observer. Retaining $L$ coefficients, indexed from $k=0$ through $L-1$, gives $L=K_{\mathrm{ISI}}=H_{\mathrm{ISI}}+1$, where $K_{\mathrm{ISI}}$ is the first omitted coefficient index and $H_{\mathrm{ISI}}$ is the number of retained past symbols. When $K_{\mathrm{ISI}}$ is not fixed for a prescribed comparison, the simulator chooses it using a tail-mass criterion,
\begin{equation}
  \sum_{k=K_{\mathrm{ISI}}}^{\infty} h_{k,m}^{(X)} \leq \epsilon_{\mathrm{ISI}}h_{0,m}^{(X)},
  \label{eq:isi_tail_criterion}
\end{equation}
with an implementation cap in long-tail regimes. The reference scenario uses $\epsilon_{\mathrm{ISI}}=10^{-3}$ and $K_{\mathrm{ISI}}\leq60$. This concentration-domain criterion bounds omitted kernel mass relative to the current-symbol coefficient. Matched simulations at several memory depths assess whether the selected truncation depth is adequate for the reported receiver metrics.

\subsection{Aptamer Binding and OECT Transduction}
\label{subsec:baseline_binding_oect}

The binding module maps local concentration to the number of occupied sites on each functionalized gate. For target channel $X$ and matching species $m$, the deterministic mean-field model follows Langmuir kinetics \cite{langmuir_adsorption_1918},
\begin{align}
  \frac{dN_b^{(X)}(t)}{dt}
  &=k_{\mathrm{on},m}^{\mathrm{eff}}c_{m,\mathrm{obs}}^{(X)}(t)
  \left(N_{\mathrm{apt},X}-N_b^{(X)}(t)\right) \nonumber\\
  &\quad - k_{\mathrm{off},m}N_b^{(X)}(t),
  \label{eq:langmuir_ode}
\end{align}
where $N_{\mathrm{apt},X}$ is the effective number of sites. An optional phenomenological transport-limitation index $\mathrm{Da}\geq0$ sets $k_{\mathrm{on},m}^{\mathrm{eff}}=k_{\mathrm{on},m}/(1+\mathrm{Da})$. Affinity-sensor response can be jointly limited by mass transport and binding kinetics \cite{lubken_dynamic_2022}. Unless otherwise stated, the baseline uses $\mathrm{Da}=0$. The hydrogel-matched \CTRL channel has no intended specific binding in the reference scenario. Equations~\eqref{eq:transport_convolution}--\eqref{eq:langmuir_ode} use a passive, nondepleting concentration field, so binding neither removes molecules from nor feeds back on transport. The peak occupancy-to-emission ratio $\eta_{\mathrm{occ}}=\max_t N_b(t)/\widetilde{N}_m$ is reported as an emitted-count consistency diagnostic rather than as capture efficiency. The nominal case is paired with a signal-matched low-occupancy case using $N_{\mathrm{apt}}/5$ and $5N_m$. For an isolated single release with occupancy reset, $N_b(t)\leq\widetilde{N}_m$ is a necessary but insufficient condition for mass conservation. All reported current traces and SER results use this passive field.

For stochastic simulations of $N_{\mathrm{apt}}$ independent equivalent sites, \eqref{eq:langmuir_ode} is advanced over each interval by an exact two-state transition under piecewise-constant concentration. At sample $j$ with step size $\Delta t$, the occupancy update is
\begin{equation}
  N_{b,j}=N_{b,j-1}+U_{\mathrm{on},j}-U_{\mathrm{off},j},
  \label{eq:stochastic_binding_update}
\end{equation}
with conditionally independent endpoint-transition counts
\begin{align}
  U_{\mathrm{on},j} &\sim \operatorname{Binomial}\left(N_{\mathrm{apt}}-N_{b,j-1},p_{\mathrm{on},j}\right), \nonumber\\
  U_{\mathrm{off},j} &\sim \operatorname{Binomial}\left(N_{b,j-1},p_{\mathrm{off},j}\right).
  \label{eq:stochastic_binding_binomial}
\end{align}
For interval $(t_{j-1},t_j]$, the implementation holds $c_j=c_{m,\mathrm{obs}}^{(X)}(t_j)$ constant and defines $\lambda_j=k_{\mathrm{on}}^{\mathrm{eff}}c_j$, $\mu=k_{\mathrm{off}}$, and $\phi_j=1-\exp[-(\lambda_j+\mu)\Delta t]$. The transition probabilities are $p_{\mathrm{on},j}=\lambda_j\phi_j/(\lambda_j+\mu)$ and $p_{\mathrm{off},j}=\mu\phi_j/(\lambda_j+\mu)$, with continuous zero-rate limits. For a reversible target channel at constant $c$ with $k_{\mathrm{on}}^{\mathrm{eff}}c+k_{\mathrm{off}}>0$, the equilibrium is
\begin{align}
  \theta_{\infty}&=\frac{k_{\mathrm{on}}^{\mathrm{eff}}c}
  {k_{\mathrm{on}}^{\mathrm{eff}}c+k_{\mathrm{off}}}, \nonumber\\
  \mathbb{E}[N_b]&=N_{\mathrm{apt}}\theta_{\infty}, \nonumber\\
  \operatorname{Var}(N_b)&=N_{\mathrm{apt}}\theta_{\infty}(1-\theta_{\infty}).
  \label{eq:binding_equilibrium}
\end{align}

Occupancy-reset (no-ISI) benchmarks retain only the current-symbol concentration and begin every symbol with all receptor sites unbound, $N_b=0$. ISI-enabled sequences superpose prior concentration responses and propagate terminal occupancy. The parameter $K_{\mathrm{ISI}}$ sets the first omitted concentration tap, while receptor state remains continuous.

The OECT module maps occupied sites to current through a quasi-static small-signal model using transconductance and volumetric capacitance parameters \cite{khodagholy_high_transconductance_2013,rivnay_thickness_2015,rivnay_oect_review_2018,friedlein_device_physics_2018}. This symbol-window approximation requires OECT and readout relaxation times short relative to $W$. A dynamic, nonlinear, or bias-dependent model can replace it through the same module interface. For channel $X$,
\begin{equation}
  i_{\mathrm{sig},X}(t)=g_m\frac{q_{\mathrm{eff},X}e}{C_{\mathrm{tot}}}N_b^{(X)}(t),
  \label{eq:oect_transduction}
\end{equation}
where $g_m$ is the transconductance, $C_{\mathrm{tot}}$ is the effective gate/channel capacitance, $e$ is the elementary charge, and $q_{\mathrm{eff},X}$ is a signed dimensionless lumped coupling coefficient. The physical effective charge factor entering the voltage perturbation is therefore $q_{\mathrm{eff},X}e$. The sign of $q_{\mathrm{eff},X}$ is part of the receiver configuration and can be changed between scenarios. The baseline tri-channel case uses opposite signs for \DA and \HT, motivated by opposite-signed dopamine and serotonin responses reported for n-type $\mathrm{In_2O_3}$ aptamer field-effect transistors \cite{nakatsuka_aptamer_fet_2018}. The \CTRL channel uses $q_{\mathrm{eff},\mathrm{CTRL}}=0$.

The detector input is the baseline-removed drain-current perturbation
\begin{equation}
  i_X(t)=i_{\mathrm{sig},X}(t)+i_{\mathrm{n},X}(t).
  \label{eq:total_current}
\end{equation}
After the nominal direct current (DC) and static readout offset are removed, $I_{\mathrm{DC}}$ appears only as the operating-point scale of the low-frequency PSD terms below.
The noise module uses an explicit one-sided current-noise PSD in \si{\ampere\squared\per\hertz}. It includes thermal, flicker, and drift-like components,
\begin{align}
  S_{\mathrm{th}}(f)&=\frac{4k_BT}{R_{\mathrm{ch}}}, \label{eq:thermal_psd}\\
  S_{1/f}(f)&=\frac{K_f I_{\mathrm{DC}}^2}{\max(f,f_{\ell})},\quad K_f=\frac{\alpha_H}{N_c}, \label{eq:flicker_psd}\\
  S_{\mathrm{drift}}(f)&=\frac{K_d I_{\mathrm{DC}}^2}{\max(f,f_{\ell})^2}, \label{eq:drift_psd}
\end{align}
Here $k_B$ is Boltzmann's constant, $T$ is temperature, $R_{\mathrm{ch}}$ is channel resistance, $\alpha_H$ is the Hooge coefficient, $N_c$ is the effective carrier count, $K_d$ is the phenomenological drift coefficient, and $f_{\ell}$ is a fixed low-frequency plateau corner independent of record length. Equation~\eqref{eq:thermal_psd} follows Johnson--Nyquist noise and \eqref{eq:flicker_psd} uses a Hooge-style parameterization \cite{johnson_thermal_1928,nyquist_thermal_1928,hooge_1f_1969}. Low-frequency OECT noise motivates the explicit flicker and drift terms \cite{stoop_charge_noise_2017}. Thermal, flicker, and drift component processes are mutually independent. Each component $\ell\in\{\mathrm{th},1/f,\mathrm{drift}\}$ is assigned a symmetric positive-semidefinite correlation matrix $\mathbf{R}_{\ell}$ with unit diagonal. Its cross-spectrum is
\begin{equation}
  S_{XY,\ell}(f)=R_{XY,\ell}\sqrt{S_{X,\ell}(f)S_{Y,\ell}(f)}.
  \label{eq:noise_correlation_matrix}
\end{equation}
This full-matrix form separates DA--\CTRL, 5-HT--\CTRL, and DA--5-HT correlation. An equicorrelation model is one admissible special case. The electrical grid has sample rate $f_{s,e}\geq2B_{\mathrm{det}}$ and is independent of the molecular transport step $\Delta t$. One multichannel record is synthesized over the complete calibration or evaluation sequence and is then partitioned into symbol windows. Colored processes therefore remain continuous across symbol boundaries. The real-valued fast Fourier transform (rFFT) DC coefficient is retained as a zero-mean stochastic low-frequency draw evaluated on the $f_{\ell}$ plateau with the half-bin endpoint weight of the one-sided PSD integral. It is distinct from the deterministic baseline current and offset removed before detection. Calibration and evaluation use independent electrical-noise records and therefore independent DC realizations. The Nyquist endpoint is treated analogously. For singular positive-semidefinite component correlation matrices, the square-root factor from an eigendecomposition transforms independent draws into correlated noise.

For a rectangular integration window of duration $W$, define $H_W(f)=W\operatorname{sinc}(fW)$ using the normalized sinc convention. Here $B_{\mathrm{det}}$ is the one-sided detection bandwidth and $\mathbf{S}_{\ell}^{(1)}(f)$ is the one-sided multichannel cross-spectral-density matrix. The component charge-covariance matrices are
\begin{equation}
  \mathbf{\Sigma}_{Q,\ell}=\int_{0}^{B_{\mathrm{det}}}
  \mathbf{S}_{\ell}^{(1)}(f)\lvert H_W(f)\rvert^2df,
  \qquad \mathbf{\Sigma}_{Q}=\sum_{\ell}\mathbf{\Sigma}_{Q,\ell}.
  \label{eq:charge_noise_covariance}
\end{equation}
White thermal noise therefore contributes $S_{\mathrm{th}}\int_0^{B_{\mathrm{det}}}\lvert H_W(f)\rvert^2df$, which explicitly accounts for the finite detection bandwidth. The same one-sided convention is used for both the analytical normalizer and sampled noise.

\subsection{Decision Statistics and Receiver Diagnostics}
\label{subsec:baseline_detection_metrics}

The detector first converts each channel current to a charge statistic by integrating over a tail-anchored decision window,
\begin{equation}
  q_i^{(X)}=\int_{T_s-W}^{T_s}i_X(t_i+t)dt,\qquad t_i=(i-1)T_s.
  \label{eq:charge_statistic}
\end{equation}
Thus, the detector operates on electrical charge-domain statistics rather than directly on local concentration.
For amplitude decisions, the matched control statistic is subtracted from the selective-channel statistic,
\begin{equation}
  \widetilde{q}_i^{(t)}=q_i^{(t)}-g_cq_i^{(\mathrm{CTRL})},\qquad t\in\{\mathrm{DA},\mathrm{5\text{-}HT}\},
  \label{eq:control_referenced_charge}
\end{equation}
where $g_c=1$ for the matched baseline and denotes the effective \CTRL subtraction gain varied in the control-gain mismatch experiment. With channel order DA, 5-HT, and \CTRL, the referenced covariance is $\mathbf{A}_c\mathbf{\Sigma}_Q\mathbf{A}_c^{\mathsf T}$ for
\begin{equation}
  \mathbf{A}_c=\begin{bmatrix}1&0&-g_c\\0&1&-g_c\end{bmatrix}.
  \label{eq:control_covariance_transform}
\end{equation}
Writing $\mathbf{\Sigma}_{c}=\mathbf{A}_c\mathbf{\Sigma}_Q\mathbf{A}_c^{\mathsf T}$, the residual correlation used by the CSK branch is $\rho_{cc}=(\mathbf{\Sigma}_{c})_{12}/\sqrt{(\mathbf{\Sigma}_{c})_{11}(\mathbf{\Sigma}_{c})_{22}}$. It is therefore derived from the specified spectra, component correlation matrices, integration window, bandwidth, and $g_c$, rather than imposed independently.
The baseline MoSK detector uses unreferenced selective-channel charges and forms the polarity-corrected identity statistic
\begin{equation}
  \Gamma_{\mathrm{MoSK},i}=\frac{\operatorname{sgn}(q_{\mathrm{eff},\mathrm{DA}})q_i^{(\mathrm{DA})}-
  \operatorname{sgn}(q_{\mathrm{eff},\mathrm{5\text{-}HT}})q_i^{(\mathrm{5\text{-}HT})}}{\sigma_{\Delta}},
  \label{eq:mosk_statistic}
\end{equation}
where $\sigma_{\Delta}^2=\mathbf{w}_{\mathrm{M}}^{\mathsf T}\mathbf{\Sigma}_Q\mathbf{w}_{\mathrm{M}}$ and $\mathbf{w}_{\mathrm{M}}=[\operatorname{sgn}(q_{\mathrm{eff},\mathrm{DA}}),-\operatorname{sgn}(q_{\mathrm{eff},\mathrm{5\text{-}HT}}),0]^{\mathsf T}$. This quadratic form is important because positively correlated noise adds, rather than cancels, when the configured effective-charge signs make $\mathbf{w}_{\mathrm{M}}=[-1,-1,0]^{\mathsf T}$. The CSK and Hybrid amplitude branches use control-referenced statistics. For CSK-4 on target axis $t$ with the other selective axis $o$, the baseline dual-channel statistic is
\begin{equation}
  \Gamma_{\mathrm{CSK},i}=\frac{\widetilde{q}_i^{(t)}}{\sigma_t}-\rho_{cc}\frac{\widetilde{q}_i^{(o)}}{\sigma_o},
  \label{eq:csk_statistic}
\end{equation}
where $\sigma_t^2=(\mathbf{\Sigma}_c)_{tt}$ and $\sigma_o^2=(\mathbf{\Sigma}_c)_{oo}$. Under an ideal target-axis mean shift and equal class covariances given by the modeled electrical covariance $\mathbf{\Sigma}_c$, this statistic is proportional to the Gaussian discriminant. Release and binding variability enter stochastic calibration rather than $\mathbf{\Sigma}_c$. The Hybrid detector first estimates the molecule-identity bit using \eqref{eq:mosk_statistic}. It then thresholds the selected-axis statistic
\begin{equation}
  \Gamma_{\mathrm{Hyb},i}=\frac{\widetilde{q}_i^{(t)}}{\sigma_t},\qquad
  t=\begin{cases}
    \mathrm{DA}, & \widehat{m}_i=\mathrm{DA},\\
    \mathrm{5\text{-}HT}, & \widehat{m}_i=\mathrm{5\text{-}HT}.
  \end{cases}
  \label{eq:hybrid_statistic}
\end{equation}
Thresholds are estimated from stochastic calibration sequences and locked during evaluation. For adjacent Gaussian-like classes, the threshold is placed at the equal-density boundary when a monotone root lies within the class interval and at the midpoint otherwise. At each molecule budget, one budget-specific 128-symbol stochastic calibration realization is used in two passes to fit the Hybrid identity and amplitude thresholds. Its seed is disjoint from the four 200-symbol adaptive-search records and the twenty 500-symbol held-out records. Operating-point-specific calibration is retained as an optimistic front-end benchmark. A separate transfer study compares operating-point-specific, nominal fixed, and scheduled-refresh calibration with a single midpoint refresh across nearby operating points that vary molecule budget, distance, and effective \CTRL subtraction gain.

For a sequence of $N_{\mathrm{sym}}$ symbols, the empirical SER is
\begin{equation}
  \widehat{P}_{\mathrm{e}}=\frac{1}{N_{\mathrm{sym}}}\sum_{i=1}^{N_{\mathrm{sym}}}\mathbf{1}_{\widehat{b}_i\neq b_i}.
  \label{eq:empirical_ser}
\end{equation}
For a fixed receiver scenario, let $\widehat{\boldsymbol{\tau}}_{N_m}$ denote the threshold set fitted at budget $N_m$ and locked before evaluation. On the search-selected grid $\mathcal{G}$, $N_{m,\varepsilon}^{\mathcal{G}}$ is the smallest $N_m\in\mathcal{G}$ for which $P_{\mathrm{e}}(N_m\mid\widehat{\boldsymbol{\tau}}_{N_m})\leq\varepsilon$. This quantity is conditional on the tested grid, fitted thresholds, and stated scenario. It is not averaged over search or calibration randomness. The held-out procedure reports whether the evaluated points meet the SER criterion and, when supported, a one-sided bound on this quantity.
Finite-sample uncertainty is reported using the central 95\% Wilson interval for a binomial proportion with normal quantile $z=1.96$ \cite{wilson_probable_1927}. For observed error fraction $\widehat{p}$ and sample size $n$, the interval center and half-width are
\begin{align}
  c_W&=\frac{\widehat{p}+z^2/(2n)}{1+z^2/n},\nonumber\\
  \Delta_W&=\frac{z}{1+z^2/n}\sqrt{\frac{\widehat{p}(1-\widehat{p})}{n}+\frac{z^2}{4n^2}}.
  \label{eq:wilson_interval}
\end{align}
The Wilson upper endpoint is $U_W=c_W+\Delta_W$, clipped to the probability interval. Low-frequency noise and ISI can correlate adjacent errors, so sequence comparisons use a moving-block bootstrap stratified by independent simulation seed \cite{kunsch_block_bootstrap_1989}. Blocks remain within seed records, and final sampled blocks are truncated to the seed-record length. For the stochastic and ISI comparisons, the block length is the maximum of 16 symbols, the ceiling of an initial-positive-sequence autocorrelation-time estimate, and $H_{\mathrm{ISI}}+1$. The tested-grid budget, calibration-transfer, control-correlation, and configuration-reuse comparisons use 16-symbol blocks. Sensitivity of the held-out budget result is also evaluated with block lengths 8 and 32. Central 95\% percentile intervals use 1000 replicates for receiver comparisons and 2000 for the held-out budget evaluation.

Let $U_{\mathrm{MB}}$ be the bootstrap upper endpoint and $U^{\star}=\max(U_W,U_{\mathrm{MB}})$ the prespecified SER upper limit used by the decision rule. The adaptive search selects $N_{\mathrm{search}}$. An independent held-out seed set then evaluates that fixed candidate and its lower searched neighbor $N_{-}$. A tested point meets the criterion when $U^{\star}\leq\varepsilon$. If $U^{\star}>\varepsilon$, the held-out data are insufficient to show that the point meets the criterion, and no replacement candidate is selected. If both evaluated points meet the criterion, the data support $N_{m,\varepsilon}^{\mathcal{G}}\leq N_{-}$. The search, calibration, and held-out seed sets are mutually disjoint. The primary bootstrap conditions on the observed seed strata. A separately reported sensitivity uses 10,000 replicates that resample seed records and then noncircular length-16 blocks within each selected record. The held-out budget evaluation uses no ISI, operating-point-specific calibration, and $\varepsilon=10^{-2}$. For transmitted symbol $B\in\mathcal{A}$ and decoded $\widehat{B}$, empirical mutual information is computed from their confusion matrix,
\begin{equation}
  \widehat{I}_{\mathrm{emp}}(B,\widehat{B})=\sum_{b,\widehat{b}}\widehat{p}_{b,\widehat{b}}
  \log_2\!\frac{\widehat{p}_{b,\widehat{b}}}{\widehat{p}_b\widehat{p}_{\widehat{b}}},
  \label{eq:empirical_mi}
\end{equation}
where zero-probability terms are omitted. This finite-sample plug-in statistic quantifies decoded-symbol information retention under the prescribed input distribution and decoder. Under the standard independent-sample approximation for a fully occupied four-by-four confusion matrix, its nominal leading positive-bias scale is $(4-1)^2/(2N_{\mathrm{sym}}\ln 2)$, or $6.49\times10^{-3}$ bits/symbol at $N_{\mathrm{sym}}=1000$. We report it only for matched settings, with block-bootstrap intervals and normalization by $\log_2|\mathcal{A}|$ when modes differ. SER and the held-out tested-grid molecule-budget bound remain the primary reliability measures.

\subsection{Implementation Verification Protocol}
\label{subsec:validation_ladder}

The verification protocol combines equation and unit checks, independent numerical references, and convergence tests. Table~\ref{tab:validation_ladder} reports the discrepancies and corresponding numerical tolerances. The stochastic tolerances are numerical-verification criteria rather than confidence limits. Wider component-specific tolerances accommodate finite-record variation and unstable relative errors for weak covariance terms. Together, raw discrepancies, bootstrap coverage or standardized discrepancies, and downstream convergence characterize numerical agreement.

\begin{table*}[!t]
\caption[Numerical-verification checks]{Numerical-verification checks for the baseline implementation. Point-kernel and finite-area observer checks are reported separately.}
\label{tab:validation_ladder}
\centering
\footnotesize
\renewcommand{\arraystretch}{1.08}
\begin{tabularx}{\textwidth}{p{0.20\textwidth} p{0.28\textwidth} X}
\toprule
\textbf{Check group} & \textbf{Independent or analytical reference} & \textbf{Observed discrepancy and numerical tolerance} \\
\midrule
Point transport and finite release & Closed-form point Green's function and independent rectangular-burst quadrature & Instantaneous point-kernel error $0$ and maximum finite-release relative error $4.51\times10^{-12}$. Limits are $5\times10^{-12}$ and $2\times10^{-5}$, respectively. \\
Binding & Langmuir moments and independent continuous-time Gillespie simulation for a reduced 8000-site, \SI{120}{\nano\molar} benchmark \cite{gillespie_exact_1977} & Exact-transition transient-mean normalized root-mean-square error (NRMSE) $3.99\times10^{-4}$ against $0.02$, maximum SSA $|z|=2.10$ against $4.5$, and exact-transition and SSA final-variance errors $0.0504$ and $0.0464$ against the three-standard-error limit $0.137$. \\
OECT and sequence noise & Static gain, one-sided PSD covariance integral, and sampled component covariance & Static-gain relative error $1.41\times10^{-16}$ against $10^{-12}$. Across 198 covariance checks, median and maximum relative errors are $0.0154$ and $0.170$, below component-specific tolerances of $0.30$--$0.55$, with 188 analytical entries inside bootstrap 95\% intervals. \\
Detector and ISI components & Synthetic Gaussian detector theory and independent point-impulse ISI quadrature & Maximum detector SER error $1.19\times10^{-3}$, with every case inside its Monte Carlo precision limit. Maximum tap error $3.70\times10^{-4}$ against $0.008$. \\
Brownian mean transport & Exact three-dimensional Brownian endpoints and shell-count uncertainty & All signal points lie within $3\sigma$, exceeding the required $0.90$ fraction. Maximum median $|z|=0.814$ against $1.25$, peak error $0.00519$ against $0.08$, and point-to-shell bias $0.0156$ against $0.04$. \\
Independent end-to-end reconstruction & Point-observer reconstruction with fourth-order Runge--Kutta (RK4) binding at $\Delta t=\SI{0.25}{\second}$ & Zero decision mismatches and maximum scaled charge or detector-statistic error $1.68\times10^{-5}$ against $10^{-4}$. \\
Convergence and calibration & Time-step, retained coefficient count, sequence length, seeds, threshold length, and noise-scale checks & Maximum $h_0$ change $0.00318$ against $0.03$. For DA at $T_s=\SI{60}{\second}$ and $W=\SI{36}{\second}$, retained mass at $L=8$ is $0.9997$. All 80 calibration criteria pass, with maximum threshold scaled error $0.0447$ against $0.15$ and noise-scale error $0.188$ against $0.35$. \\
\bottomrule
\end{tabularx}
\end{table*}

Together, these checks verify implementation consistency over the tested ranges.

The point-transport, Brownian, and independent end-to-end references use the analytical point kernel independently of the finite-area quadrature. Section~\ref{subsec:results_applicability_transfer} evaluates the finite-area calculation, nodewise binding, the passive-model occupancy-to-emission ratio, and local propagation-count effects separately.

\section{Numerical Verification and Receiver Demonstrations}
\label{sec:results}

The fixed scenario is given in Table~\ref{tab:case_study_parameters}. Component checks verify implementation consistency, while the demonstrations evaluate finite-area observation, calibration transfer, a held-out molecule-budget bound, control referencing, and ISI.

\begin{table*}[!t]
\caption[Reference receiver scenario]{Reference DA/5-HT/CTRL receiver scenario. Values are assumed model parameters rather than values fitted to measured device data.}
\label{tab:case_study_parameters}
\centering
\footnotesize
\renewcommand{\arraystretch}{1.12}
\begin{tabularx}{\textwidth}{p{0.15\textwidth} p{0.43\textwidth} X}
\toprule
\textbf{Group} & \textbf{Baseline values} & \textbf{Role in the case study} \\
\midrule
Transport and timing & $r=\SI{45}{\micrometerunit}$, $D_{\DA}=4.9\times10^{-10}$ and $D_{\HT}=5.3\times10^{-10}~\si{\meter\squared\per\second}$, $\alpha=0.20$, $\lambda=1.6$, $k_{\mathrm{clear}}=0.01~\si{\per\second}$, $T_{\mathrm{rel}}=\SI{10}{\milli\second}$, $T_s=\SI{20}{\second}$, and $W=\SI{12}{\second}$ & Square-area observer with $L_g=\SI{200}{\micrometerunit}$, order-24 quadrature, and molecular step $\Delta t=\SI{0.01}{\second}$. \\
Binding and selectivity & $N_{\mathrm{apt}}=2\times10^8$ sites/gate, $k_{\mathrm{on}}=10^5~\si{\per\molar\per\second}$, $k_{\mathrm{off},\DA}=0.015~\si{\per\second}$, and $k_{\mathrm{off},\HT}=0.003~\si{\per\second}$ & Target-specific stochastic occupancy. \CTRL has no intended specific analyte-binding term. \\
OECT transduction & $g_m=\SI{5}{\milli\siemens}$, $C_{\mathrm{tot}}=\SI{50}{\nano\farad}$, $R_{\mathrm{ch}}=\SI{500}{\ohm}$, $I_{\mathrm{DC}}=\SI{100}{\micro\ampere}$, $q_{\mathrm{eff},\DA}=-0.35$, and $q_{\mathrm{eff},\HT}=+0.35$ & Quasi-static symbol-window mapping with dimensionless $q_{\mathrm{eff}}$ and physical factor $q_{\mathrm{eff}}e$ in \eqref{eq:oect_transduction}. \\
Noise and reference & $T=\SI{310}{\kelvin}$, $f_{s,e}=\SI{128}{\hertz}$, $B_{\mathrm{det}}=\SI{50}{\hertz}$, $f_{\ell}=10^{-3}~\si{\hertz}$, $\alpha_H=10^{-3}$, $N_c=4.5\times10^{11}$, and $K_d=10^{-16}~\si{\hertz}$ & The thermal correlation matrix is the identity. Baseline flicker and drift off-diagonal correlations are $0.9$. The \CTRL sweep holds DA--5-HT correlation at $0.9$ while varying the two selective--\CTRL entries together. \\
Inference and randomization & MoSK, CSK, and Hybrid detectors with one 128-symbol stochastic calibration realization per operating point, independent symbol, release, binding, and electrical-noise streams, and disjoint calibration, search, and held-out seeds & Matched receiver contrasts use $1000$ symbols. Adaptive molecule-budget search uses $4\times200$ symbols per point, and the held-out budget evaluation uses $20\times500$ symbols. The target SER is $10^{-2}$. \\
\bottomrule
\end{tabularx}
\end{table*}

The effective site count, charge-coupling factors, drift coefficient, affinity parameters, and noise-component correlation matrices are assumed model parameters rather than measurements from a fabricated tri-channel receiver.

\subsection{Numerical Verification and Convergence}
\label{subsec:results_validation}

All checks in Table~\ref{tab:validation_ladder} pass. The Brownian benchmark uses exact three-dimensional endpoints, shell counts, and binomial count uncertainty across six distance and clearance cases. Every signal point lies within $3\sigma$ of the shell reference, with maximum median $|z|=0.814$, peak error $0.519\%$, and point-to-shell bias $1.56\%$.

The binding check uses 960 exact-transition and 960 independent continuous-time Gillespie realizations of an 8000-site receptor at \SI{120}{\nano\molar}. The binomial fourth moment gives a $4.57\%$ relative Monte Carlo standard error for each final sample variance. Final-variance deviations for the exact-transition method and SSA are $5.04\%$ and $4.64\%$, both below the prespecified three-standard-error limit of $13.7\%$. Final-mean errors are $0.0292\%$ and $0.0102\%$, while maximum SSA transient-mean $|z|$ is $2.10$. The deterministic reconstruction independently joins point transport, RK4 binding, OECT charge, subtraction, and decisions. It gives zero decision mismatches and maximum scaled statistic error $1.68\times10^{-5}$.

\begin{figure*}[!tb]
  \centering
  \includegraphics[width=0.90\textwidth]{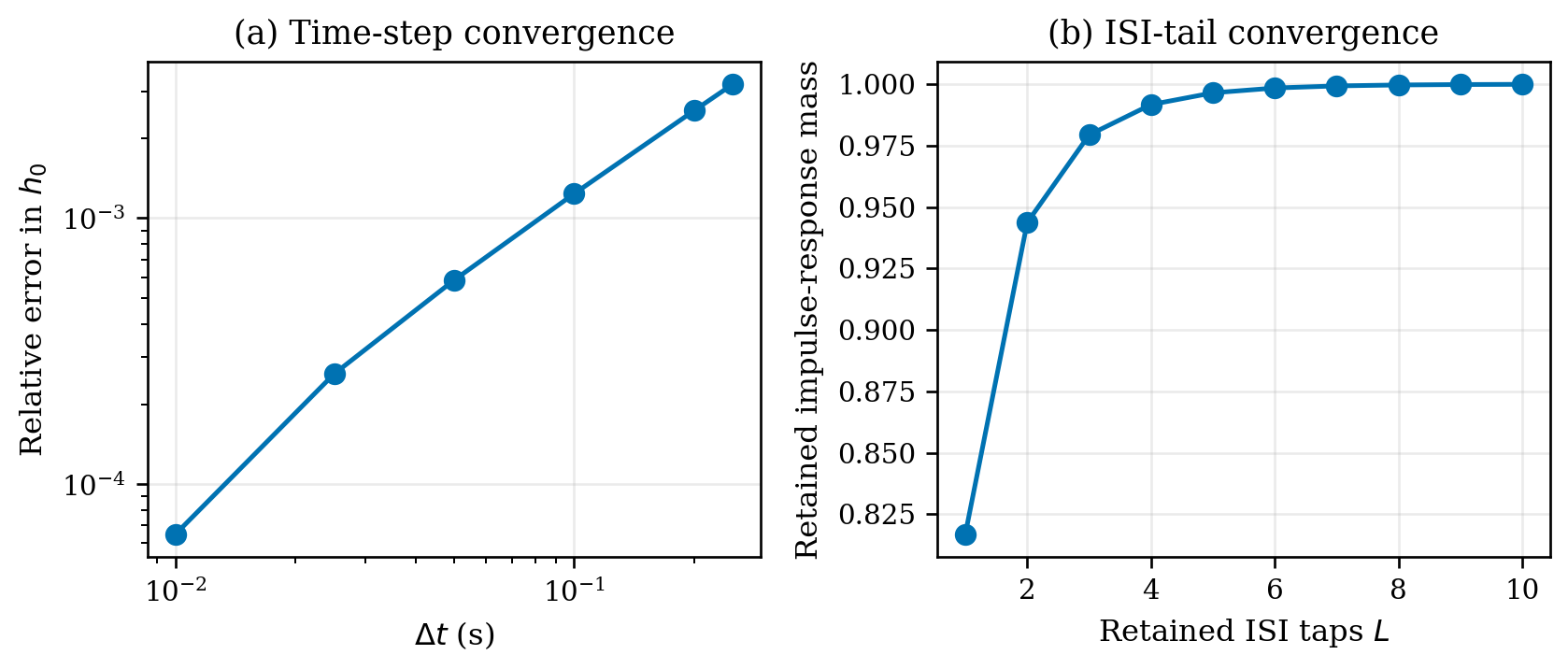}
  \caption[Numerical convergence]{Numerical convergence. (a) First-tap relative error against a $\SI{0.005}{\second}$ time-step reference. (b) DA retained mass versus $L$ at $T_s=\SI{60}{\second}$ and $W=\SI{36}{\second}$, referenced to $k=0,\ldots,60$.}
  \label{fig:convergence_results}
\end{figure*}

The point-kernel check spans $r\in\{25,45,90\}~\si{\micrometerunit}$ and $k_{\mathrm{clear}}\in\{0,0.01\}~\si{\per\second}$, with finite-release checks at $r\in\{25,45\}~\si{\micrometerunit}$ and $k_{\mathrm{clear}}=0.01~\si{\per\second}$. In Fig.~\ref{fig:convergence_results}, first-tap error falls from $0.318\%$ at $\Delta t=\SI{0.25}{\second}$ to $0.00650\%$ at the adopted $\SI{0.01}{\second}$, relative to $\SI{0.005}{\second}$. DA retained mass rises from $0.944$ at $L=2$ to $0.9997$ at $L=8$. Sequence length changes SER by $3.13\times10^{-4}$, below the reference Wilson half-width $8.56\times10^{-4}$. All 80 calibration criteria pass. Maximum scaled errors are $0.0447$ against $0.15$ for stochastic thresholds and $0.188$ against $0.35$ for calibration-noise scale.

Machine-specific measurements under Windows 11 Pro on a 13th Gen Intel Core i9-13950HX (24 physical cores and 32 logical processors) with \SI{96}{\giga\byte} RAM, Python 3.11.9, and NumPy 2.4.4 gave a median runtime of $281.06~\si{\second}$ over three runs (interquartile range $2.34~\si{\second}$) and a peak process resident set size of $297~\si{\mega\byte}$ for a 400-symbol stochastic Hybrid sequence at $\Delta t=\SI{0.01}{\second}$.

A separate deterministic MoSK workload at $\Delta t=\SI{0.25}{\second}$ and $H_{\mathrm{ISI}}=3$ required median runtimes $(4.61,9.43,18.59)~\si{\second}$ for $(200,400,800)$ symbols, with six repetitions per length. Its three-point log--log slope was $1.0055$ ($R^2=0.9998$), indicating approximately linear measured scaling over this range. This controlled workload is not a timing surrogate for the stochastic Hybrid case.

\subsection{Sequence-Wide Noise and Covariance}
\label{subsec:results_noise}

\begin{figure*}[!tb]
  \centering
  \includegraphics[width=0.95\textwidth]{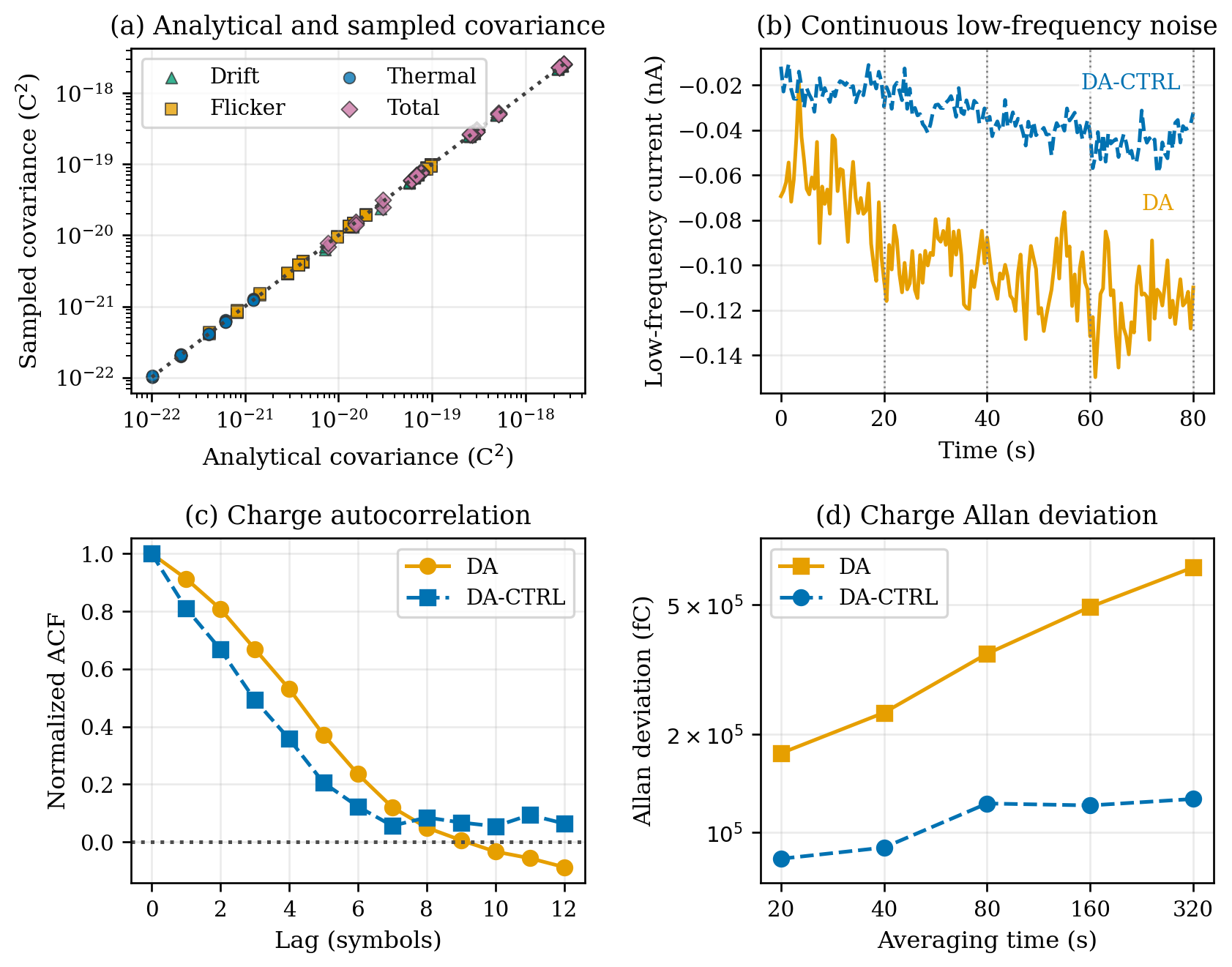}
  \caption[Sequence-wide noise characterization]{Electrical-noise characterization: (a) agreement between analytical and sampled charge covariance, (b) a continuous low-frequency trace, (c) charge autocorrelation, and (d) \CTRL-referenced Allan deviation.}
  \label{fig:noise_evidence}
\end{figure*}

Figure~\ref{fig:noise_evidence} checks $f_{s,e}\in\{128,256\}~\si{\hertz}$, $B_{\mathrm{det}}\in\{25,50\}~\si{\hertz}$, $T_s\in\{20,60\}~\si{\second}$, and $W\in\{6,12,36\}~\si{\second}$. All 198 covariance checks meet their component-specific tolerances. Covariance errors have median $1.54\%$ and maximum $17.0\%$, 188 analytical entries lie within record-bootstrap 95\% intervals, and maximum $|z|$ is $2.83$. Across 99 sample-rate pairs, relative differences between sampled-to-analytical covariance ratios have median $2.03\%$ and maximum $26.9\%$, while the maximum absolute paired difference is $0.224$. The worst case is the total cross-covariance between the two \CTRL-referenced selective channels at $B_{\mathrm{det}}=\SI{50}{\hertz}$, $T_s=\SI{20}{\second}$, and $W=\SI{12}{\second}$. Its ratios are $0.833$ and $1.056$, with $|z|=2.35$ and $0.758$, both below the prespecified limit of 3. Because \eqref{eq:charge_noise_covariance} is analytically sample-rate invariant, the paired comparison measures discretization sensitivity. Numerical agreement is evaluated separately at each sample rate.

In the displayed realization, subtraction changes lag-one charge autocorrelation from $0.914$ to $0.811$ and Allan deviation at $\SI{320}{\second}$ from $6.51\times10^{-10}$ to $1.27\times10^{-10}~\si{\coulomb}$. These illustrative values complement the quantitative comparison of analytical and sampled covariance.

\subsection{Finite-Area Observation and Calibration Transfer}
\label{subsec:results_applicability_transfer}

\begin{figure*}[!tb]
  \centering
  \includegraphics[width=0.98\textwidth]{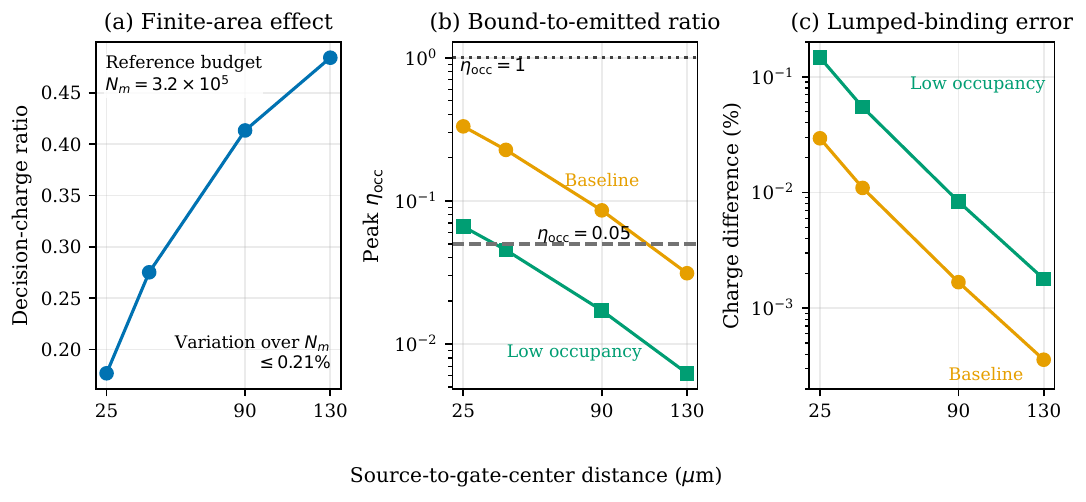}
  \caption[Finite-area observation and binding]{Effects of finite-area observation and the binding approximation at the modeled \SI{200}{\micrometerunit} square gate. (a) Ratio of area-averaged to center-point decision-charge magnitude. Variation over the tested molecule-budget grid is at most $0.21\%$. (b) Occupancy-to-emission ratio. (c) Nodewise-to-lumped decision-charge difference.}
  \label{fig:geometry_applicability}
\end{figure*}

Figure~\ref{fig:geometry_applicability} shows that replacing center-point observation with square-area averaging has a substantial effect. At $r=\SI{45}{\micrometerunit}$ and $N_m=3.2\times10^5$, the square-area observer retains $0.275$ of the center-point decision-charge magnitude and changes $\eta_{\mathrm{occ}}$ from $0.713$ to $0.227$. Solving binding at each quadrature node in this DA geometry check changes decision charge by only $0.0109\%$ in the baseline and $0.0546\%$ in the signal-matched low-occupancy case. At the baseline test point $N_m=2.5\times10^5$, the $(N_{\mathrm{apt}}/5,5N_m)$ parameter set retains $0.9991$ and $0.9992$ of baseline \DA and \HT charge while reducing $\eta_{\mathrm{occ}}$ to $0.0454$ and $0.0501$, respectively. Order-24 area quadrature agrees with order 32 within $3.23\times10^{-7}$ for the tested summaries.

A DA-only particle simulation with 32 realizations per observation depth assesses local propagation-count fluctuations. Across prism thicknesses of $1$--$20~\si{\micrometerunit}$, its conditional decision-charge coefficient of variation is $0.202$--$0.328\%$. Particle-mean charge differs by at most $0.072\%$ from the continuum result integrated over the same finite observation prism. Together with the observer and binding comparisons, this supports the continuum mean approximation over the tested high-count regime while keeping the finite-geometry and passive-observer assumptions explicit.

\begin{figure*}[!tb]
  \centering
  \includegraphics[width=0.96\textwidth]{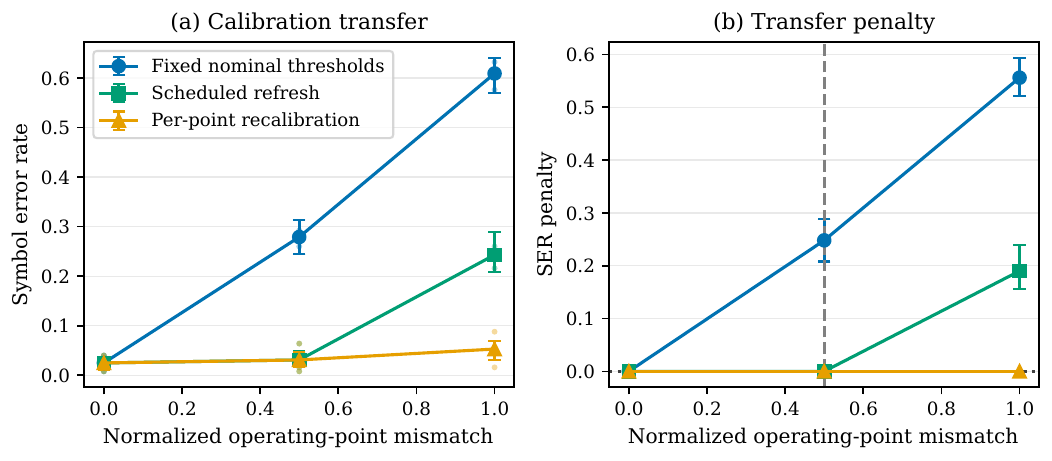}
  \caption[Calibration-transfer experiment]{Calibration transfer as molecule budget, distance, and \CTRL gain change jointly. The nominal point is $(N_m,r,g_c)=(2.5\times10^5,\SI{45}{\micrometerunit},1)$ and the endpoint is $(2.0\times10^5,\SI{55}{\micrometerunit},0.85)$. Fixed nominal thresholds, scheduled refresh, and per-point recalibration are evaluated using common evaluation seeds that are disjoint from calibration. Points show seed-level estimates. Error bars are seed-stratified moving-block intervals. Penalties are paired relative to per-point recalibration, which serves as the optimistic reference.}
  \label{fig:calibration_transfer}
\end{figure*}

In Fig.~\ref{fig:calibration_transfer}, all policies give SER $0.025$ nominally. At the midpoint, fixed nominal thresholds give $0.279$, while scheduled refresh and per-point recalibration both give $0.031$. At the endpoint, per-point recalibration gives $0.053$, while fixed and scheduled thresholds give $0.609$ and $0.243$. Their paired penalties are $0.556$ (95\% block interval $[0.522,0.594]$) and $0.190$ ($[0.156,0.240]$). Because budget, distance, and effective \CTRL gain change together, this experiment evaluates calibration transfer under joint mismatch rather than attributing the loss to one factor.

\subsection{Independent Receiver Evaluation}
\label{subsec:results_case_study}

\begin{figure*}[!tb]
  \centering
  \includegraphics[width=0.98\textwidth]{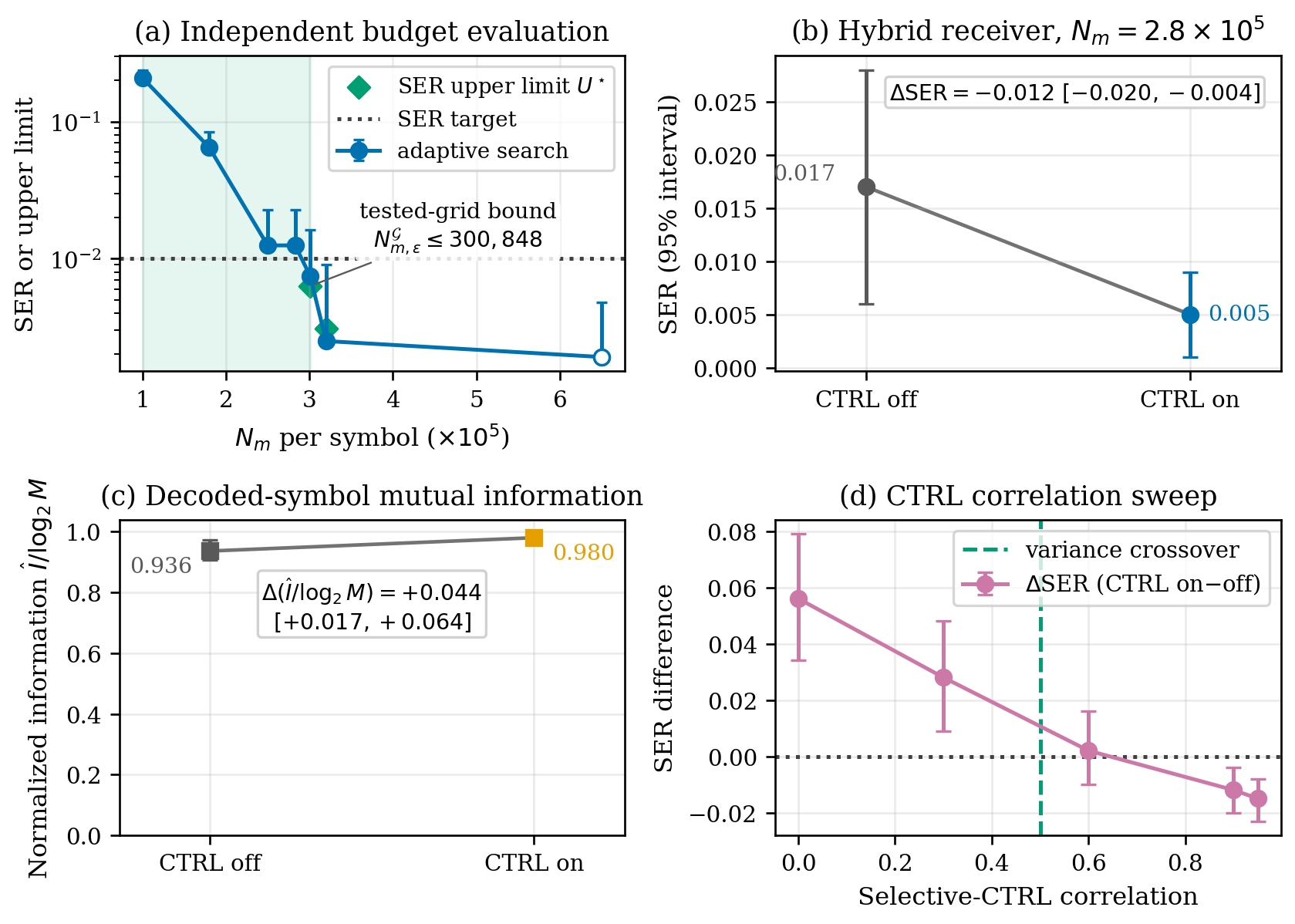}
  \caption[Independent receiver evaluation]{Tri-channel receiver evaluation. (a) Adaptive search for Hybrid detection with \CTRL, followed by independent evaluation at the selected budget and its lower neighbor. The passive-field test resets receptor occupancy, excludes ISI, and calibrates each operating point. Diamonds denote $U^{\star}$, and shading marks the tested-grid bound. Panels (b) and (c) show SER and normalized empirical mutual information under the same conditions. (d) \CTRL-on minus \CTRL-off SER as selective--\CTRL correlation varies with DA--5-HT correlation fixed. Negative values favor \CTRL. Search, calibration, and evaluation use disjoint seed sets.}
  \label{fig:case_study_results}
\end{figure*}

The adaptive search selects $N_{\mathrm{search}}=320{,}000$ molecules/symbol and the lower neighbor $N_{-}=300{,}848$. Twenty 500-symbol held-out records per point produce 19 and 45 errors. With 16-symbol bootstrap blocks, $U^{\star}=0.0031$ and $0.0063$. Both points continue to meet the criterion at block lengths 8 and 32. Secondary 95\% seed-plus-block upper endpoints are $0.0036$ and $0.0068$. All lie below $\varepsilon=0.01$, supporting $N_{m,\varepsilon}^{\mathcal{G}}\leq300{,}848$ molecules/symbol, approximately $0.50~\si{\atto\mole}$, conditional on the tested grid, passive observation, occupancy reset, and operating-point-specific calibration.

With symbols, seeds, physical setting, and calibration matched, Fig.~\ref{fig:case_study_results}(b) and (c) vary only the \CTRL policy. At $N_m=2.8\times10^5$, SER changes $0.017\to0.005$, paired $-0.012$ ($[-0.020,-0.004]$), while normalized empirical mutual~information changes $0.936\to0.980$, paired $0.044$ ($[0.017,0.064]$). The unnormalized estimates are $1.873$ and $1.960$ bits/symbol.

With DA--5-HT correlation fixed at $0.9$, selective--\CTRL correlations $(0,0.6,0.9)$ give paired SER changes of $0.056$ ($[0.034,0.079]$), $0.002$ ($[-0.010,0.016]$), and $-0.012$ ($[-0.020,-0.004]$). The sweep varies the configured low-frequency component correlation. For target $t\in\{\mathrm{DA},\mathrm{5\text{-}HT}\}$, \CTRL subtraction changes integrated charge variance by $g_c^2(\mathbf{\Sigma}_{Q})_{\mathrm{CTRL},\mathrm{CTRL}}-2g_c(\mathbf{\Sigma}_{Q})_{t,\mathrm{CTRL}}$. Its zero gives the crossover near $0.5$, while detection and calibration also affect SER.

At $N_m=(2.5,4.0,6.5)\times10^5$, single-axis DA CSK-4 yields SERs $(0.245,0.078,0.011)$ and empirical mutual information $(1.043,1.568,1.914)$ bits/symbol. Four 250-symbol records per point use disjoint 128-symbol calibration. The SER intervals at the first and third budgets are $[0.202,0.280]$ and $[0.002,0.025]$. The single-axis case uses the same module interfaces and receiver metrics. A direct performance comparison or molecule-budget bound for this configuration would require a matched protocol.

\begin{figure*}[!tb]
  \centering
  \includegraphics[width=0.98\textwidth]{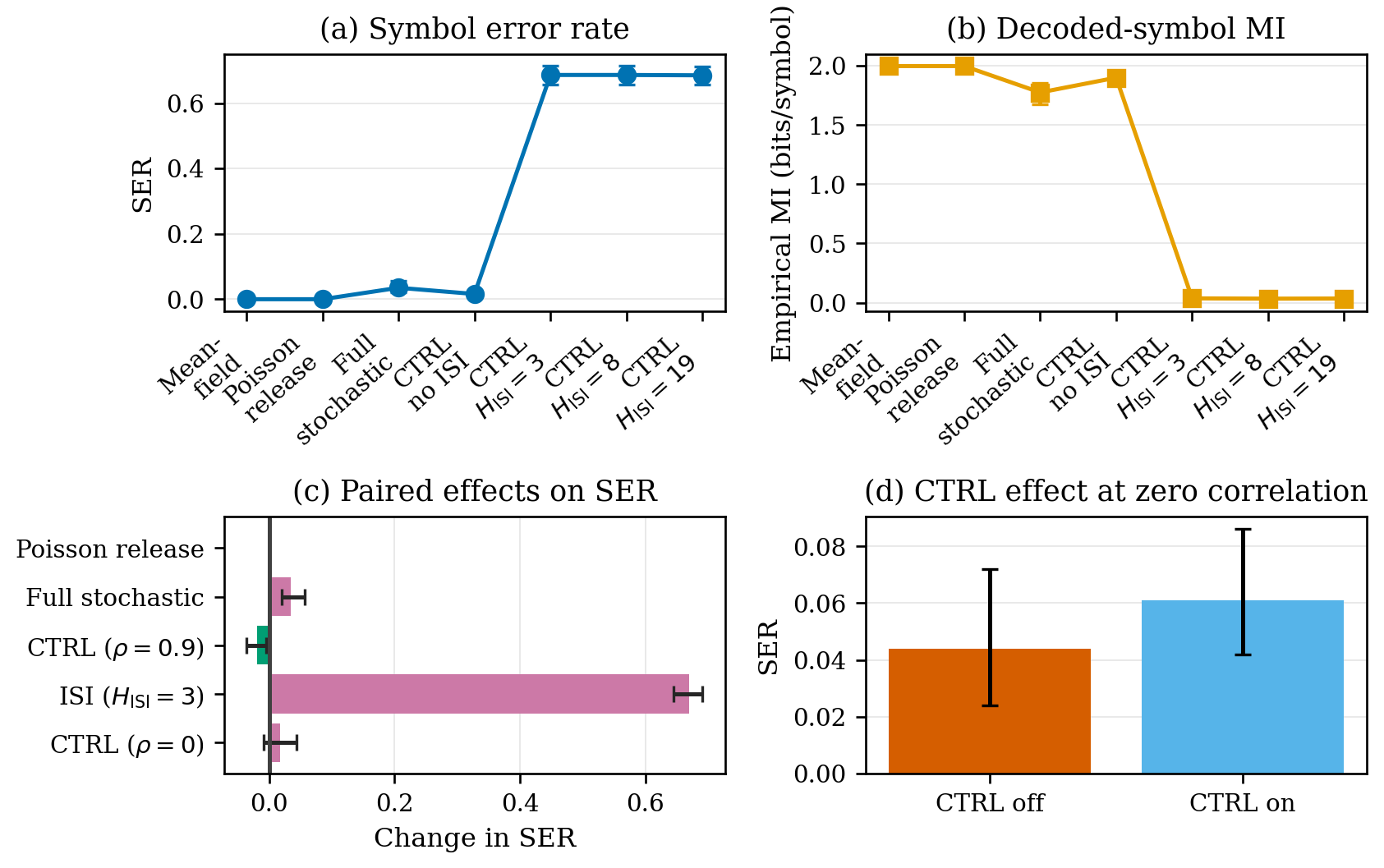}
  \caption[Stochastic and ISI tests]{Stochastic and ISI tests at $N_m=2.5\times10^5$ under a common operating point. Panels (a) and (b) compare mean-field, Poisson-release, full-stochastic, \CTRL no-ISI, and $H_{\mathrm{ISI}}\in\{3,8,19\}$ cases. Each case is calibrated separately using common random streams, with calibration and evaluation seeds disjoint. SER error bars show the conservative envelope of Wilson and seed-stratified moving-block intervals, while empirical mutual information uses seed-stratified moving-block intervals. (c) Paired SER changes as individual mechanisms are introduced. (d) \CTRL effect at zero selective--\CTRL correlation.}
  \label{fig:realism_ladder}
\end{figure*}

\subsection{Stochastic and ISI Tests}
\label{subsec:results_realism}

Figure~\ref{fig:realism_ladder} reports zero errors in 1000 mean-field or Poisson-only symbols (Wilson upper endpoint $0.00383$). Including all modeled receiver stochasticity increases SER by $0.035$ ($[0.020,0.057]$).

At high and zero selective--\CTRL correlation, enabling \CTRL changes SER from $0.035$ to $0.016$ and from $0.044$ to $0.061$, respectively. The paired changes are $-0.019$ ($[-0.036,-0.005]$) and $0.017$ ($[-0.009,0.043]$). The latter interval includes zero.

With high-correlation \CTRL and $T_s=\SI{20}{\second}$, $H_{\mathrm{ISI}}=3$ gives SER $0.686$ ($[0.657,0.714]$), a no-ISI penalty of $0.670$ ($[0.645,0.691]$), and empirical mutual information $0.0360$ versus $1.899$ bits/symbol. Separately calibrated depths 8 and 19 give SERs $0.686$ ($[0.657,0.714]$) and $0.685$ ($[0.656,0.713]$), with successive paired changes $0.000$ ($[-0.007,0.007]$) and $-0.001$ ($[-0.006,0.004]$). At $H_{\mathrm{ISI}}=19$, the minimum retained-mass fraction is $0.99942$. Its omitted-tail ratio is $9.74\times10^{-4}$ versus $1.27\times10^{-3}$ at the preceding candidate, so 19 is the smallest depth meeting the $10^{-3}$ concentration-tail rule. Empirical mutual information remains $0.0334$--$0.0360$ bits/symbol across these depths. Matched random streams support paired comparisons across memory depths.

The signal-matched low-occupancy case shows the same trends. At high correlation, the no-\CTRL and \CTRL SERs are $0.039$ ($[0.020,0.066]$) and $0.013$ ($[0.006,0.023]$), compared with $0.046$ ($[0.026,0.075]$) and $0.088$ ($[0.062,0.117]$) at zero correlation. With $H_{\mathrm{ISI}}=3$, high-correlation \CTRL reaches $0.696$ ($[0.676,0.717]$), confirming that low no-ISI SER alone does not establish sequence reliability.

\section{Discussion and Limitations}
\label{sec:discussion_limitations}

OECT-MC receiver comparisons should account explicitly for geometry, covariance, calibration, and memory. For the reference case, center-point decision charge is about 3.6 times the finite-gate value. \CTRL helps only with favorable covariance, scheduled refresh reduces transfer loss, and retained memory yields high SER. One end-to-end curve conflates these effects.

\frameworkname separates them using explicit module interfaces, a common charge-domain statistic, and matched protocols. Geometry and binding shape signal charge. Covariance determines whether \CTRL helps, calibration controls threshold transfer, and transport and receptor dynamics govern memory-dependent performance. A fixed downstream gain rescales charge but does not reproduce the distance-dependent finite-area effect. \CTRL cannot remove selective-channel ISI, and per-point calibration cannot evaluate transfer.

Numerical checks confirm equation, implementation, and interface consistency over the tested ranges. Device prediction requires measurements of transport, binding, OECT response, multichannel noise, control mismatch, and aging. At the nominal point, the $27.5\%$ area-to-point charge ratio reflects passive spatial averaging, not capture efficiency. Evaluating capture and depletion requires mass-conserving reactive transport. The held-out tested-grid bound is conditional on the passive-observer model, occupancy reset, no ISI, and per-point calibration.

Because DA and 5-HT dissociation times exceed $T_s=\SI{20}{\second}$, occupancy reset isolates no-ISI behavior but does not represent a memoryless receptor. The finite-area observer, nodewise binding model, and single-axis case show that the shared pipeline accommodates model and configuration changes.

\FloatBarrier

\section{Conclusion}
\label{sec:conclusion}

\frameworkname links molecular release to calibrated OECT charge decisions through explicit module interfaces. It provides finite-area observation, full-sequence multichannel colored-noise synthesis on an independent electrical grid, charge-covariance analysis, independent evaluation, and block-bootstrap intervals. Using the reference case, it separates geometry, covariance, calibration, and memory effects across charge, SER, and decoded-symbol information. Its held-out tested-grid bound is conditional on the passive-observer model, occupancy reset, no ISI, and per-point calibration. Device prediction requires measurements and mass-conserving reactive transport when capture or depletion matters. \frameworkname supports controlled, reproducible comparison of OECT-MC receiver hypotheses and systematic exploration of candidate receiver designs.

\FloatBarrier

\bibliographystyle{IEEEtran}
\bibliography{references}

\ifincludeauthorbios
\begin{IEEEbiography}[{\biophoto{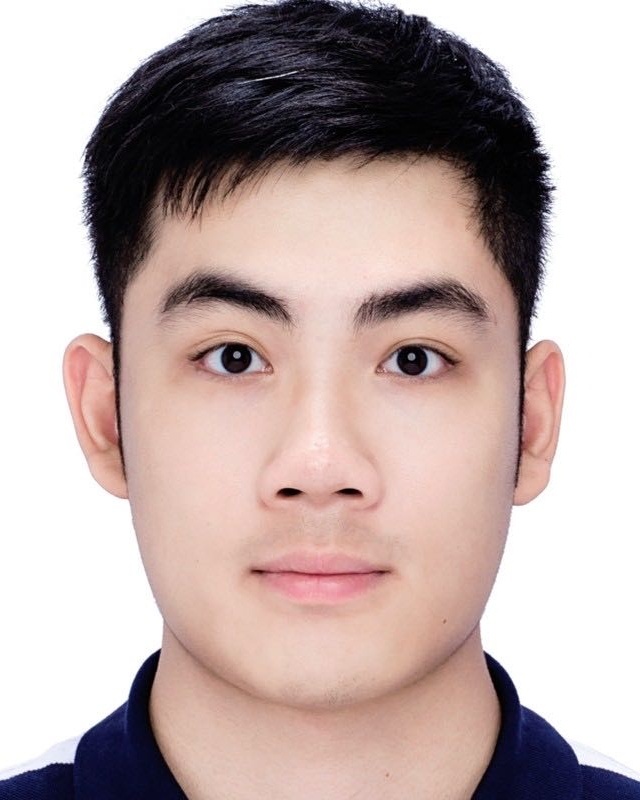}}]{Hongbin Ni}
\hangafter=-10\relax
(Student Member, IEEE) received the B.Eng.\ degree in Mechatronic Engineering from The University of Manchester, U.K., in 2021 and the M.Sc.\ degree in Biomedical Engineering from Imperial College London, U.K., in 2022. He is currently pursuing the Ph.D.\ degree in engineering with the Centre for neXt Communications, Department of Engineering, University of Cambridge, U.K. His research interests include molecular communication and its applications to neural interfaces.
\end{IEEEbiography}

\begin{IEEEbiography}[{\biophoto{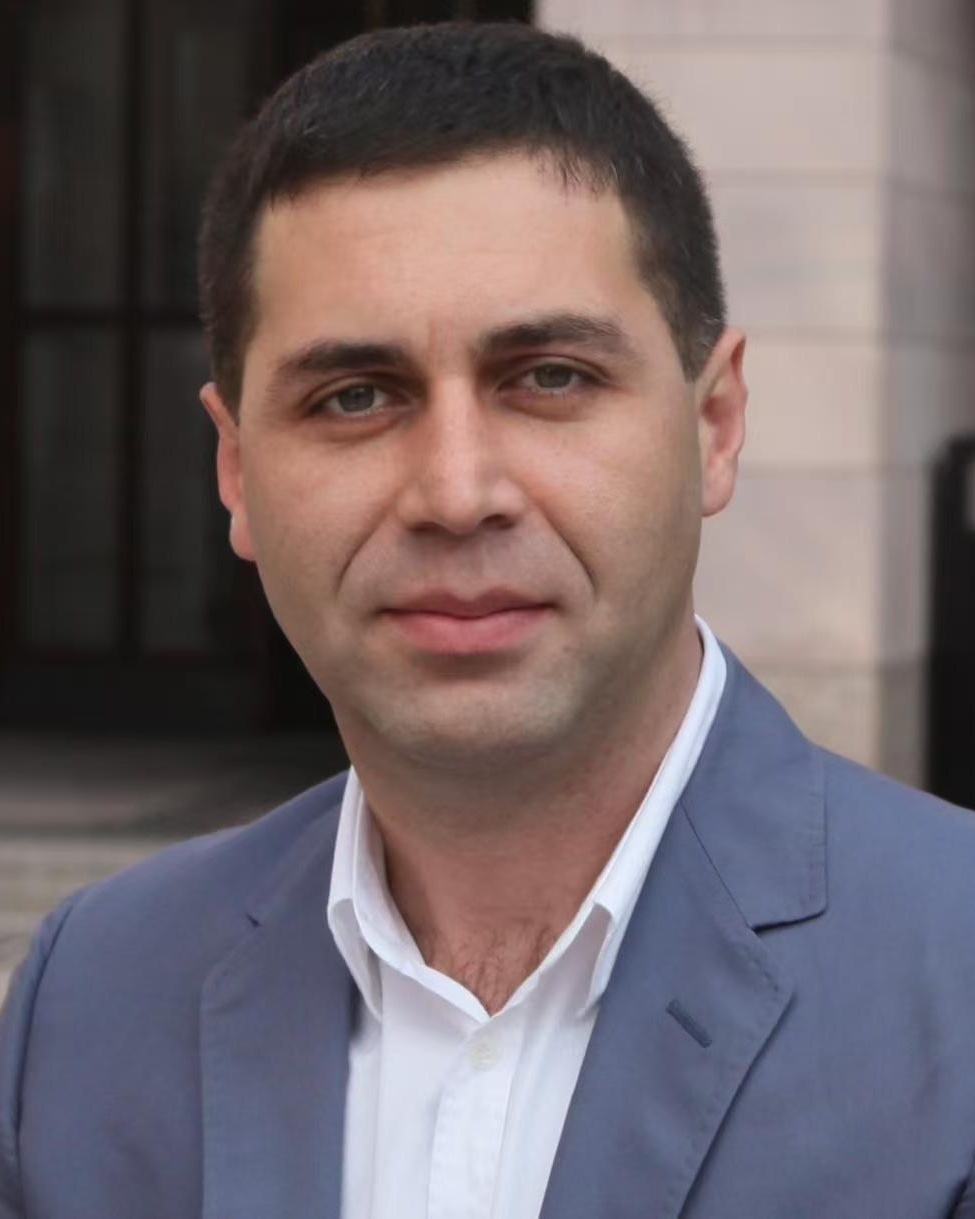}}]{Ozgur B.~Akan}
(Fellow, IEEE) received the Ph.D. degree in electrical and computer engineering from the Georgia Institute of Technology, U.S.A., in 2004. He is a Professor with the University of Cambridge, U.K., where he leads the Centre for neXt Communications, and is Director of CXC at Ko\c{c} University, T\"{u}rkiye. His interests include wireless, space, quantum, and molecular communications.
\end{IEEEbiography}

\vfill
\fi

\end{document}